# Amplification due to Two-Stream Instability of Self-Electric and Magnetic Fields of an Ion Beam Propagating in Background Plasma

Erinc K. Tokluoglu, Igor D. Kaganovich, Johan A. Carlsson, Kentaro Hara, Edward A. Startsev

*Princeton Plasma Physics Laboratory, Princeton, New Jersey 08543, USA*

Propagation of charged particle beams in background plasma as a method of space charge neutralization has been shown to achieve a high degree of charge and current neutralization and therefore enables nearly ballistic propagation and focusing of charged particle beams. Correspondingly, use of plasmas for propagation of charged particle beams has important applications for transport and focusing of intense particle beams in inertial fusion and high energy density laboratory plasma physics. However, the streaming of beam ions through a background plasma can lead to development of the two-stream instability between the beam ions and the plasma electrons. The beam electric and magnetic fields enhanced by the two-stream instability can lead to defocusing of the ion beam. Using particle-in-cell (PIC) simulations, we study the scaling of the instability-driven self-electromagnetic fields and consequent defocusing forces with the background plasma density and beam ion mass. We identify plasma parameters where the defocusing forces can be reduced.

## I-Introduction

Beam-plasma systems have a wide range of applications in Inertial Confinement Fusion (ICF) [1,2,3,4], including fast ignition fusion for ICF [5,6,7], high energy density physics [8,9,10], astrophysics [11,12,13], atomic physics [14] and basic physical phenomena [15,16,17]. The background plasma presents a means of current and charge neutralization for charged particle beams, enabling the ballistic propagation of the intense beam pulse [18,19,20,21,22,23,24,25,26,27,28]. However, the beam streaming through the background plasma can lead to the development of many different instabilities [29, 30], from which the two-stream instability is the fastest [29].

*This paper reviews past theoretical and experimental studies of effects of the two-stream instability on ion beam propagation in background plasma. We also investigate scaling of deleterious effects of two stream instability on the ion beam propagation in background plasma with the beam and plasma parameters and determine optimum conditions where these deleterious effects can be minimized.*

The theory of the nonlinear development of the two-stream instability for the ion beam pulse propagating in plasma was developed initially in Ref. [23]. In the case of an electron beam propagating in background plasma, the current driven by the two-stream instability can result in enhancement of the total current (beam and plasma return currents) and self-magnetic field. This was demonstrated using a PIC code [25] and was experimentally observed in Refs. [26, 28]. In the case of an ion beam propagating in background plasma, it has been shown that the two-stream instability between beam ions and electrons can also strongly affect the electron return current in the plasma [23, 25, 27, 31, 32]. In the case of an intense ion beam, the non-linear time-averaged current driven by the two-stream instability between the beam ions and the plasma electrons has been shown to reverse the direction of the total current and the self-magnetic field. Because of the self-magnetic field reversal, in presence of two-stream instability, the self-magnetic field causes defocusing of the ion beam. Note that in absence of two-stream instability the self-magnetic field causes focusing of the ion beam [33].

The electric field of plasma waves driven by the two-stream instability can set up a time-averaged radial electric field (similar to ponderomotive field). This radial electric field defocuses the propagating ion beam [32] similarly to the radial electric field produced by electron pressure.

Therefore, in the case of an ion beam propagating in background plasma in the presence of the two-stream instability both the radial electric field and the reversal of the self-magnetic field act together to defocus the beam as it propagates. Furthermore, high intensity longitudinal electric fields of plasma waves generated by two-stream instability can strongly modulate the density of propagating ion beam longitudinally – produce bunching of initially long beam pulses [34,35,36, 37].

However, for short propagation in the plasma, the deleterious effects of the instabilities were shown to be small experimentally and theoretically for the Neutralized Drift Compression eXperiments (NDCX): NDCX-I [20-22, 32] and NDCX-II experiments [38, 39]. The established level of self-electromagnetic fields in the presence of the ion-beam driven two–stream instability strongly depends on the instability saturation mechanism [32, 37]. It has been shown both in the presence and absence of external magnetic field, that ion-beam-driven two-stream instability saturates by the wave-particle trapping of either of beam ions or plasma electrons depending on the beam and plasma parameters: the beam density, plasma density and ion mass [27, 32, 37]. Therefore, these parameters determine the peak self-electric and magnetic fields produced by the instability.

Moreover, the value of the ratio of beam radius relative to the plasma skin depth also influences both the self-electric and self-magnetic fields. For NDCX-II parameters it was shown previously [32, 34, 37] that the two-stream instability does not significantly distort the beam radial profile during propagation and compression. A diagnostic approach tracking the spot size of an extracted beamlet of small radius was proposed in Ref. [34] to detect the presence of the instability in the case of a Li$^+$ beam for NDCX-II experiment [38, 39].

We have performed a range of particle-in-cell (PIC) simulations using the PIC code LSP [40] to investigate general scaling of saturation mechanisms of two-stream instability, and corresponding self-electromagnetic fields and the defocusing force acting on beam ions as a function of background plasma density. In the first part of the paper, a proton beam with beam velocity $v_b = c/2$, where c is the speed of light in vacuum, propagating in a cold carbon plasma has been simulated. We have varied the ratio of beam density, $n_b$, and plasma density, $n_p$ and investigated the magnitudes of the self- electromagnetic fields at instability saturation and have determined the defocusing forces that the intense beam pulse experiences during propagation.

We have also studied the instability saturation mechanism for different beam ion masses and verified previously proposed scaling for two-stream instability saturation and defocusing force with the ion mass and beam and plasma density [32, 37].

Finally, for the NDCX-II Li$^+$ ion beam, we have also performed simulations in the two limits: the beam radius, $r_b$, large or small relative to the plasma skin depth, $\delta_{skin}$, $r_b > \delta_{skin}$ and $r_b \ll \delta_{skin}$. These results demonstrate the effect that the ratio of the beam radius to the plasma skin depth has on the self-electromagnetic fields and, consequently, defocusing of the ion beam pulse during propagation in the background plasma.

This paper is organized as follows. In Sec. II we give a brief description of the simulation set-up and the particle-in-cell code LSP used for the simulations. In Sec. III, we present a theoretical description of the non-linear self-electromagnetic fields generated by the ion beam driven two-stream instability and how the magnitude of self-electromagnetic fields depends on the beam and plasma parameters, including the ratio of beam density to plasma density, the beam ion mass and the beam radius. In Sect. IV we compare results of the PIC simulations with an analytical model for instability-driven electron current and discuss scaling of the self-electric and self-magnetic field at the saturation of the two-stream instability with the ratio of the beam and plasma densities. In Sect. V we demonstrate the dependence of the radial defocusing force on the beam ion mass and present the simulation results for the Li$^+$ ion beam pulse of the NDCX-II experiment and the potential implications for future heavy-ion fusion experiments. In Sect. VI we discuss the two-stream instability for a flat-top beam radial profile, which is more relevant than a Gaussian for intense beams. In Sect. VII we present simulation results of two-stream instability for a proton beam generated by a laser. Finally in Sec. VIII, we summarize key results from the previous sections and draw conclusions.

## II- Simulation Setup

As a base case, an intense proton (H$^+$) beam pulse is injected into plasma; the beam pulse radial and longitudinal profile is a Gaussian

$$n_b(r,z) = n_{b0}\exp\left(-\frac{r^2}{r_b^2} - \frac{z^2}{l_b^2}\right).$$

Here, the maximum beam density, $n_{b0} = 2 \times 10^{10}\ cm^{-3}$, the beam radius, $r_b = 2\ cm$, the beam velocity, $v_b = c/2$, where $c$ is the speed of light in vacuum, characteristic pulse duration, $t_{pulse} = l_b/v_b = 4.44\ ns$, and the total pulse duration was limited to 12 ns, (that is: no beam ions are present in the pulse outside this duration). The background plasma density, $n_p$ was varied in the simulations to investigate its' effect on self-electromagnetic fields of the ion beam pulse propagating in the background plasmas and further details are given in Sec. IV. The simulations were performed using particle-in-cell code LSP [40]. Collisions are not taken into account; therefore any effects leading to the radial beam expansion observed in the simulations are solely due to collisionless processes of ion beam defocusing by the self-electromagnetic fields.

We have simulated beam propagation in a 2D slab or cylindrical geometry. In the following, x denotes the transverse or radial direction and z denotes the direction of beam propagation; 3D velocity space was resolved with y denoting the azimuthal direction for cylindrical geometry. The field solver used for the simulation is implicit and electromagnetic with the time step, $\Delta t$, chosen to give an acceptable dispersion error and to resolve the plasma frequency, $\Delta t \ll 1/\omega_{pe}$, where $\omega_{pe}$ is the electron plasma frequency of the background plasma. The axial grid size, $\Delta z$, is chosen to satisfy $\Delta z \sim 1/(30 k_z)$, where $k_z = \omega_{pe}/v_b$ is the resonant wave number of the plasma waves. This fine resolution is needed to fully resolve the axial structure of the plasma waves, which are excited by the beam. The radial grid size is fixed at $\Delta x = 0.1\ cm$. This spatial grid provides sufficient resolution to observe changes in the beam radius and the radial displacement of beam ions. The domain size for the simulation is x: [-11, 11] cm and z: [0, 240] cm. We also employ the moving-frame algorithm. Initially the ion beam pulse is simulated in the laboratory frame, as soon as the center of the ion beam pulse reaches the center of the simulation domain (which takes 15.5 ns), the moving frame algorithm starts and the beam pulse remains always in the center of the moving frame.

To study the scaling of the self-electromagnetic field with the beam mass, we have simulated several cases for a fixed plasma density but varied beam mass. For each simulation, the axial grid size and the time step have been adjusted following the same procedure described previously. In addition, we have varied the plasma and beam parameters to modify the ratio of beam radius and skin depth. This ratio affects generation of the return current and, correspondingly, the self-electromagnetic field of the ion beam pulse propagating in the background plasma.

## III-Theoretical Overview of the Beam Self-Electromagnetic Fields in the Presence of Two-Stream Instability

An ion beam propagating in background plasma may induce the two-stream instability. For a cold plasma and beam in the one dimensional limit, the initial instability growth can be determined from the local dispersion function $D(k, \omega)$ [1]

$$D(k,\omega) = 1 - \frac{\omega_{pe}^2}{\omega^2} - \frac{\omega_b^2}{(\omega - k v_b)^2} = 0 \qquad (1)$$

Here, $\omega$, $\omega_{pe}$, $\omega_b$, $k$ and $v_b$ are the mode frequency, the background electron plasma frequency, the beam ion plasma density, the wave number and the axial directed beam velocity, respectively. Solving for the complex roots of Eq. (1) yields the oscillation frequency and the growth rate of the instability [41]. In the limit where the plasma density is significantly larger than the beam density, $n_p \gg n_b$, the maximum growth rate of instability for the resonant wave number, $k = \omega_{pe}/v_b$, is $\gamma = 0.7(\omega_{pe}\omega_b^2)^{1/3}$, where $\omega_{pe} = \sqrt{4\pi n_p e^2/m_e}$ is the electron plasma frequency, $e$ denotes unit charge, and $m_e$ is the electron mass and $\omega_b = \sqrt{4\pi n_b e^2/m_i}$ is the beam ion plasma frequency, where $m_i$ denotes the beam ion mass. If the beam propagation time through the plasma, $T_{transit}$, is sufficiently long, $\gamma T_{transit} \gg 1$, the two-stream instability develops.

The two-stream instability generates a spectrum of plasma waves, oscillating with frequencies close to the plasma frequency, $\omega_{pe}$, but with different wavelengths with a spectrum peaked around the resonant wavelength, $2\pi v_b/\omega_{pe}$ [23, 25]. The instability grows linearly from the noise starting from the beam head. Because the group velocity of the growing waves is comparable with but less than the beam velocity (in the laboratory frame of background plasma), the maximum of instability growth moves from the beam head towards the beam tail [42, 43] until instability reaches saturation due to nonlinear effects of particle trapping in the wave electric field.

There are two effects that are responsible for ion beam radial defocusing: the radial self-electric field, $E_x$, and the azimuthal self-magnetic field, $B_y$. In the presence of the beam-driven two-stream instability, large-amplitude plasma waves produce sufficiently strong axial electric field, $E_z$. Due to transverse variation in the beam profile, the axial electric field strength has a transverse gradient, $\nabla_x \langle E_z^2 \rangle$. This gradient creates a ponderomotive force in the radial direction acting on plasma electrons; this, in turn, generates a radial ambipolar electric field produced to counteract the ponderamotive force [32, 34, 37]. The ambipolar radial electric field generated this way defocuses the ion beam. The radial temporally and spatially-averaged electric field, $E_x$ is given by [32, 34, 37]

$$E_x \approx -\frac{e}{4m_e\omega_{pe}^2} \nabla_x |E_z|^2 = -\frac{1}{4e} m_e \nabla_x (v_m^e)^2. \qquad (2)$$

Here, $v_m^e$ is the amplitude of the axial electron velocity oscillation due to the instability. The radial electric field given by Eq.(2) represents a non-linear effect, because it is a quadratic function of $v_m^e$. Furthermore, because $v_m^e$ vanishes away from the beam pulse, the radial electric field is positive and is defocusing for the ion beam pulse.

The two-stream instability also significantly affects the electron return current, and, correspondingly, the azimuthal magnetic field, $B_y$. The two-stream instability provides coupling between the beam ions and plasma electrons and this effective "friction" between ion beam and electrons "drags" background plasma electrons along the beam path. Increased plasma electron flow causes increase in the electron current and yields reversal and significant enhancement of the total axial current and, consequently, the self-magnetic field [23, 32, 34, 37]. (Note that in case of an electron beam pulse instead of ion beam, the modified by two-stream instability current is in the same direction [25, 28]).

In the presence of the ion beam-driven two-stream instability, the total electron return current density can be calculated by time-averaged cross product of $<\delta n_e\ \delta v_m^e>$ of the perturbations in electron density and electron axial velocity generated by the instability [37]

$$<\delta n_e\ \delta v^e> \approx \frac{1}{2} n_p \left(\frac{v_m^e}{v_b}\right)^2 v_b. \qquad (3)$$

Here, we use the estimate for the perturbation of the electron density. From the electron continuity equation it follows that $\delta n_e \approx \delta v^e n_p k_z/\omega_{pe} \approx \delta v^e n_p/v_b$, for the plasma waves that are resonant with the beam, $k_z \approx \omega_{pe}/v_b$.

The self-magnetic field is determined by the Ampere law

$$\frac{\partial}{r\partial r} rB = \frac{4\pi e}{c}\left(n_b v_b - n_e v_{ez} - <\delta n_e\, \delta v^e>\right), \tag{4}$$

and conservation of the electron vorticity or canonical momentum. Here, $v_{ez}$ is the time-averaged over plasma wave time scale electron flow velocity. For long beam pulses so that the beam pulse length, $l_b$ is much longer than the beam radius, $r_b$, $l_b \gg r_b$, conservation of the electron vorticity gives [33]

$$eB = -\frac{\partial}{\partial r}eA_z \approx -\frac{\partial}{\partial r}cm_e v_{ez}, \tag{5}$$

where $eA_z \approx cm_e v_{ez}$ is the vector potential. Substituting Eq.(5) into Eq.(4) gives

$$-\frac{\partial}{r\partial r}r\frac{\partial}{\partial r}v_{ez} = \frac{4\pi e^2}{c^2 m_e}\left(n_b v_b - n_e v_{ez} - <\delta n_e\, \delta v^e>\right). \tag{6}$$

The second term on the right hand side describes the return current density caused by electron response to inductive electric field driven by the time dependent beam self-magnetic field [33]. If the beam radius is small compared with the skin depth, $r_b \ll c/\omega_{pe}$, the return current density can be neglected and the total current, which is the sum of the beam current $J_z^b$ and total electron current density, $J_z^e$, and can be approximately calculated using following relation [37]:

$$J_{tot} \approx J_z^b + J_z^e = J_z^b\left[1 - \frac{1}{2}\frac{n_p}{n_b}\left(\frac{v_m^e}{v_b}\right)^2\right]\quad \left(r_b \ll \frac{c}{\omega_{pe}}\right). \tag{7}$$

The second term on the right-hand side in the brackets comes from the time and space averaged term $<\delta n_e\, \delta v^e>$. It is important to note that in the limit $n_p \gg n_b$ (which is typically the case for neutralization applications), the non-linear term $\frac{1}{2}\frac{n_p}{n_b}\left(\frac{v_m^e}{v_b}\right)^2$ can exceed unity and the total current will be reversed and significantly amplified [23, 32, 34, 37].

Because current is reversed, the azimuthal self-magnetic field becomes also reversed, and the resulting $v_b \times B$ magnetic force on the beam ions becomes defocusing [37]. As discussed above the radial electric field is also defocusing, therefore both forces lead to defocusing of the ion beam pulse, and the ion beam radial profile can become significantly distorted. Figure 1 shows the beam profile evolution due to two-stream instability. Figure 1(a) shows the initial Gaussian density profile of an ion beam pulse prior to the development of the instability (the beam has traveled to the center of the simulation domain propagating in the background plasma for t = 16 ns, being injected at the boundary at t=0). Figure 1 (b) shows the same beam pulse at t = 40 ns, after nearly 6 m of propagation in plasma after the development and saturation of the two-stream instability. An increase in the beam radius and the distortion of the radial beam profile due to the non-linear defocusing forces is evident.

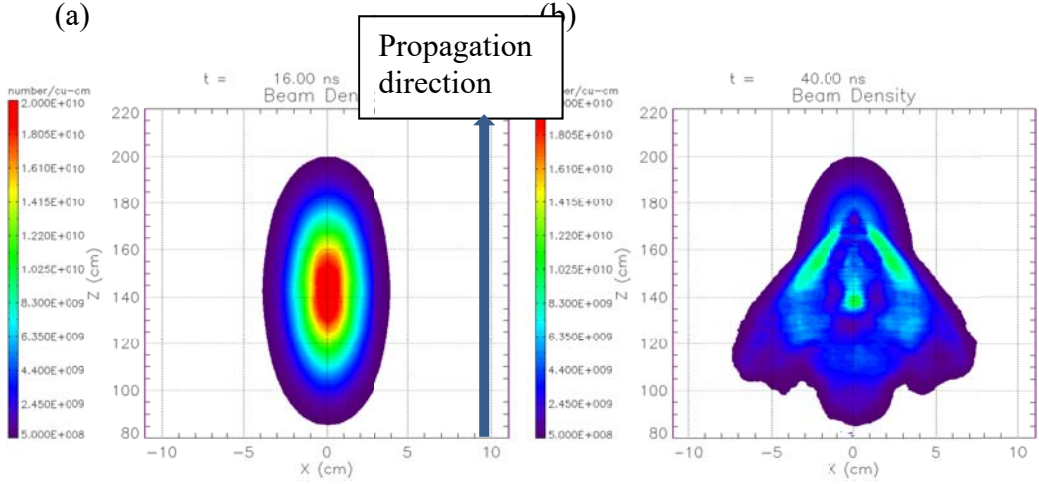

Fig.1: Colorplot of the ion beam density profile obtained in the two-dimensional PIC simulations in Cartesian coordinates; an intense proton (H$^+$) beam pulse is injected into plasma; the initial beam pulse radial and longitudinal profile is a Gaussian, $n_b(x,z) = n_{b0}\exp(-x^2/r_b^2 - z^2/l_b^2)$, where $n_p = 1.0 \times 10^{12} cm^{-3}$, $n_b = 2 \times 10^{10} \, cm^{-3}$, $r_b = 2 \, cm$, characteristic pulse duration, $t_{pulse} = l_b/v_b = 4.44 \, ns$, $v_b = c/2$ (a) prior to instability, t = 16 ns; (b) after 6 m of propagation (t = 40 ns) in plasma when instability develops and distorts the initial beam profile.

In the opposite limit, $r_b \gg c/\omega_{pe}$, the return current nearly compensates the beam current ($n_e v_{ez} \approx n_b v_b - <\delta n_e \, \delta v_m^e>$) and the remaining current can be determined from the left-hand side of equation (6), which gives

$$J_{tot} \approx J_z^b + J_z^e = -\frac{c^2}{\omega_{pe}^2} \frac{\partial}{r\partial r} r \frac{\partial}{\partial r} J_z^b \left[1 - \frac{1}{2}\frac{n_p}{n_b}\left(\frac{v_m^e}{v_b}\right)^2\right] \quad \left(r_b \gg \frac{c}{\omega_{pe}}\right). \quad (8)$$

Depending on the parameters of the beam-plasma system, the two-stream instability can saturate by two different saturation mechanisms: either due to trapping in the wave of plasma electrons or beam ions [23, 27, 32, 37]. According to Eqs. (2) and (4), the defocusing electric and magnetic forces created by the two-stream instability depend on the amplitude of axial electron velocity oscillations, $v_m^e$.

In the case of electron trapping saturation mechanism, saturation of the instability occurs when the electron oscillation amplitude reaches the phase velocity of the resonant mode [27, 32, 37]

$$v_m^e \approx \omega/k_z \approx v_b, \quad (9)$$

which is approximately equal to the beam velocity.

In the case of ion trapping, the beam ion oscillating axial velocity at saturation is given by [23, 27, 32, 37]

$$v_m^b \approx v_b - \omega/k_z \sim (\gamma/k_z) \approx (\gamma/\omega_{pe})v_b \approx (\omega_b/\omega_{pe})^{2/3} v_b, \quad (10)$$

where $\gamma$ is the growth rate of the resonant mode. The ratio of electron and ion oscillating velocities can be determined from the momentum balance, which gives

$$m_e \omega v_m^e \sim m_b(\omega - k_z v_b) v_m^b.$$

Solving for $v_m^e$ and assuming $\omega - k_z v_b \sim \gamma$ gives [27, 32, 37]

$$v_m^e \cong \left(\frac{m_b}{m_e}\right)\left(\frac{\gamma}{\omega_{pe}}\right) v_m^b \approx \left(\frac{m_b}{m_e}\right)\left(\frac{\gamma}{\omega_{pe}}\right)^2 v_b. \quad (11)$$

Depending on condition which of the species reaches their respective saturation level oscillation amplitude first, the instability saturates by the particle trapping of either beam ions or the plasma electrons. The saturation value of electron velocity oscillation amplitude normalized by beam velocity is given by [32, 37]:

$$\left(\frac{v_m^e}{v_b}\right) \sim \min[\alpha, 1], \quad \alpha \equiv \left(\frac{n_b}{n_p}\right)^{\frac{2}{3}} \left(\frac{m_b}{m_e}\right)^{1/3}. \quad (12)$$

The first limit in Eq.(12) corresponds to the case if the instability saturates by ion trapping mechanism, and the second limit corresponds to the case if the instability saturates by the electron trapping mechanism. Note that the dimensionless parameter, $\alpha$ in Eq. (12) predicts the mechanisms of two-stream instability saturation depending on the beam and plasma parameters, such as the beam density, plasma density and the beam ion mass. If

$$\alpha = \left(\frac{n_b}{n_p}\right)^{\frac{2}{3}} \left(\frac{m_b}{m_e}\right)^{1/3} > 1, \quad (13)$$

the saturation mechanism of the two-stream instability is due to electron trapping and if $\alpha < 1$, the saturation mechanism of the two-stream instability is due to the ion trapping. Therefore, for a given beam density and beam ion mass, decreasing the plasma density below some critical value, $n_{pc}$,

$$n_{pc} \equiv n_b \left(\frac{m_b}{m_e}\right)^{1/2} \quad (14)$$

results in the instability saturation by electron trapping for $n_p < n_{pc}$, in which case, the electron oscillation velocity amplitude at saturation is approximately equal to the beam velocity $v_b$. In the dense background plasma limit, $n_p > n_{pc}$, the instability saturates by the beam ion trapping saturation mechanism, and the electron velocity oscillation amplitude normalized by beam velocity is given by the scaling parameter $\alpha$ itself, according to Eq.(12).

Similarly, for fixed beam and plasma densities, for a heavier beam ion specie such that the beam ion mass is above the value given by

$$m_b > m_e \left(\frac{n_p}{n_b}\right)^{1/2}, \quad (15)$$

the two-stream instability will saturate by electrons, and for a lighter beam ion specie, the instability saturates by the trapping of beam ions. Fig. 2 presents an example of change in saturation mechanisms by decreasing the plasma density in such a manner that the scaling parameter $\alpha$ increases from 0.5 to 3.

(a)                                                                         (b)

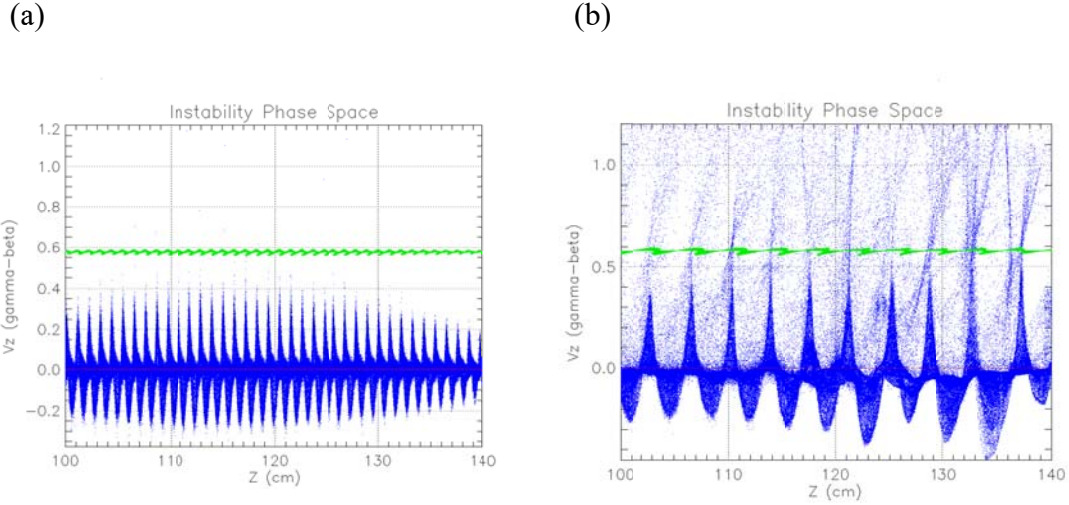

Fig.2. Phase space $P_z/m_e c = \gamma V_z/c$ vs z of electrons and ions (blue dots represent plasma electrons, red dots - plasma ions, green dots beam ions) at time of instability saturation. (a) $n_p = 2.4 \times 10^{12} cm^{-3}$, which corresponds to the ion trapping regime ($\alpha = 0.5$); (b) $n_p = 1.66 \times 10^{11} cm^{-3}$, which corresponds to the electron trapping regime ($\alpha = 3$). Beam parameters are the same as in Fig.1, the beam ion is H⁺, $n_b = 2 \times 10^{10} cm^{-3}$, $r_b = 2 cm$, $t_{pulse} = 12 ns$, $v_b = c/2$.

Figure 2 (a) shows simulated phase space for a case of a relatively high background plasma density, $n_p = 2.4 \times 10^{12}/cm^3$, corresponding to $\alpha = 0.5$. Notice that the electron oscillation amplitude is roughly half of the beam velocity as expected according to Eq.(12). Figure 2 (b) shows phase space for a lower background density, $n_p = 1.66 \times 10^{11} cm^{-3}$, corresponding to $\alpha = 3$. Because the parameter $\alpha$ is greater than unity, the electron velocity oscillation amplitude is comparable to $v_b$. A significant population of plasma electrons is trapped by the wave (distinct islands in phase space appear); this creates wave breaking and saturation of instability.

Longer evolution of the instability is studied in Ref.[44]. In this paper it is shown that a portion of the initially trapped electrons becomes detrapped and moves ahead of the ion beam pulse forming a forerunner electron beam. The self-consistent nonlinear driven turbulent state with a quasi-stationary plasma wave is set up at the head of the ion beam pulse, which lasts until the final stage when the beam ions become trapped by the plasma wave and become heated by it. The ion heating eventually extinguishes the instability. For the beam parameters discussed in that paper, these processes happen 200ns after the beam injection into the plasma, well after initial saturation of the two-stream instability when plasma wave is strongest.

## IV-Scaling of Self-Electric and Magnetic Field at Saturation of the Two-Stream Instability with Ratio of the Beam and Plasma Densities

Equation (12) shows that the ratio of ion beam density to plasma density is important in determining the saturation mechanism of the two-stream instability, and, consequently, the magnitude of the defocusing force acting on the beam ions [32, 37]. In order to study this effect we simulate interaction of the same proton beam pulse with various background plasma densities in the range $n_p \in [8 \times 10^{10}, 5 \times 10^{12}] cm^{-3}$, which corresponds to $\alpha \in [0.3, 5]$. Figure 3 shows the obtained scaling of the electron velocity oscillation amplitude and radial electric field at the saturation of the instability as a function of parameter $\alpha$.

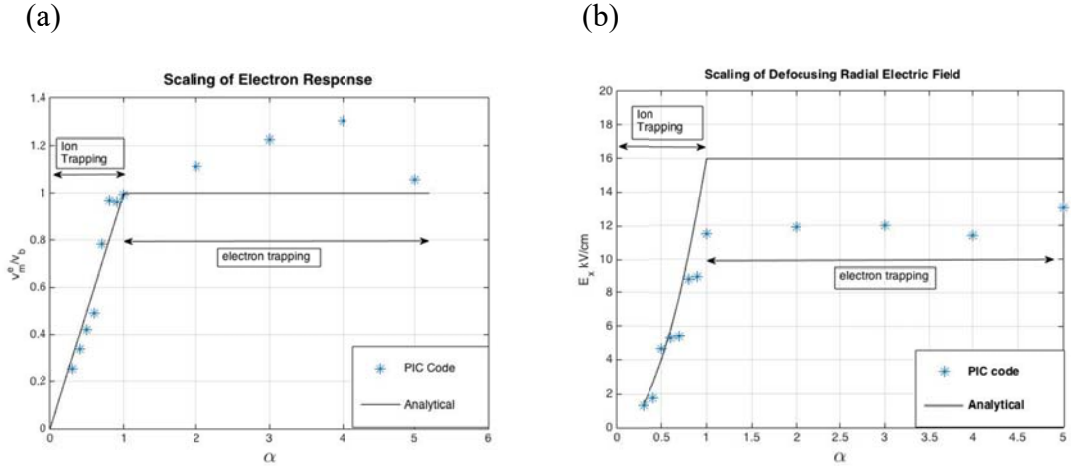

Fig.3 Scaling of (a): the electron velocity oscillation amplitude and (b): the radial defocusing electric field as a function of parameter $\alpha = C_1 (n_b/n_p)^{2/3}$, $C_1 = (m_b/m_e)^{1/3} = 12.24$ for H$^+$ beam. The beam parameters are the same as in Fig. 1. Symbols show PIC simulation results (a) the mid plane and (b) at $r \cong 1\,cm$; which corresponds to the maximum field strength; solid lines correspond to analytic estimate given by (a): Eq.(12) (b) by Eq.(2).

Figure 3 (a) demonstrates that the electron velocity oscillation amplitude at saturation behaves as described in Eq. (12). That is: if $\alpha < 1$, the axial electron velocity oscillation amplitude at saturation of instability increases linearly with $\alpha$. Once $\alpha$ becomes greater than unity, $\alpha > 1$, the two-stream instability saturates by the electron trapping mechanism, and, consequently, the electron velocity oscillation amplitude remains constant and of order of the beam velocity. Reference [44] further reports on study of long time evolution of the two-stream instability after saturation.

Figure 3 (b) shows the scaling of the radial electric field with respect to $\alpha$. Notice that in the electron trapped region ($\alpha > 1$) the radial electric field does not depend on the background density, $E_x \sim \nabla_x (v_m^e)^2 \sim v_b^2/r_b$.

Figure 4 shows the maximum value of the self-magnetic field (located at around x = 1 cm) obtained using (1) the particle-in-cell code, (2) the numerical solution of the modified Ampere's law, which includes the instability driven time-averaged non-linear electron current, Eq.(6) as well as (3) an estimate

$$B_y = \frac{2\pi n_b r_b \beta_b}{1+r_b^2 \omega_{pe}^2/c^2} \left[1 - \frac{1}{2}\frac{n_p}{n_b}\left(\frac{v_m^e}{v_b}\right)^2\right], \qquad (16)$$

which was proposed in Refs.[32,37] for a Gaussian beam profile.

The data obtained from the PIC simulations, the analytical estimate, Eq.(16), and the numerical solution for Ampere's law, Eq.(6) are in good agreement, each showing the same dependence on $\alpha$. In the transition region, where $\alpha \sim 1$ and the saturation mechanism of the instability changes, and the azimuthal magnetic field strength has a maximum as a function of $\alpha$.

Recall that in the electron trapping regime the electron velocity oscillation amplitude has a maximum and is approximately equal to the beam velocity, $v_b$. Correspondingly $B_y \approx \frac{\pi n_b r_b \beta_b}{1+r_b^2 \omega_{pe}^2/c^2}\frac{n_p}{n_b}$, and decreases with alpha due to decease of the plasma density as soon as the skin depth becomes comparable with the beam radius. Similarly in the ion trapping regime, $\alpha < 1$, the plasma density is high and the skin depth is small compared with the beam radius, correspondingly, $B_y \approx \frac{\pi n_b r_b \beta_b}{\frac{r_b^2 \omega_{pe}^2}{c^2}}\frac{n_p}{n_b}\left(\frac{v_m^e}{v_b}\right)^2 \sim \alpha^2$.

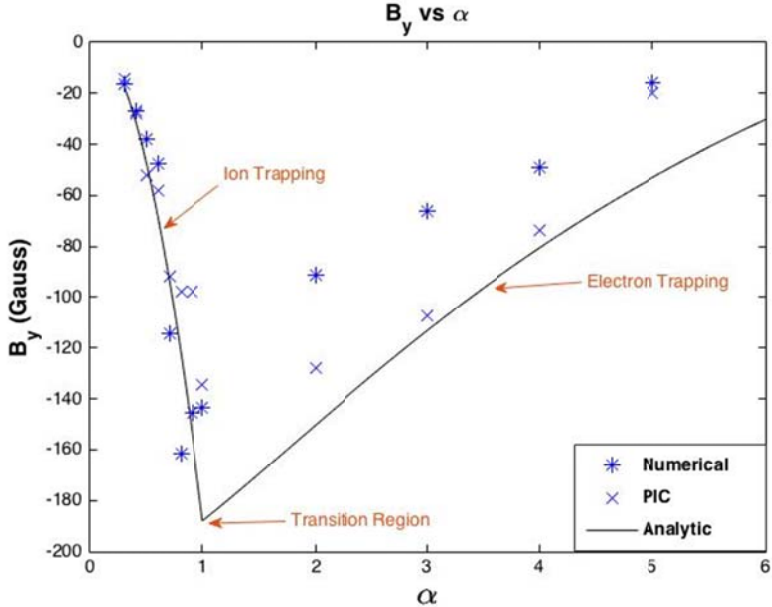

Fig.4 Maximum value of azimuthal magnetic field as a function of $\alpha$ parameter $\alpha = C_1(n_b/n_p)^{2/3}$, $C_1 = \left(\frac{m_b}{m_e}\right)^{1/3} \sim 12.24$ for H$^+$ beam. Note that the maximum magnetic field strength is attained at the transition region. Crosses correspond to the PIC simulation results, stars correspond to the solution of Eqs.(6), (8),(12), lines correspond to the analytical estimate given by Eq.(16).

The azimuthal magnetic field produced due to the two-stream instability has a defocusing effect on the ion beam, because the electric currents of ion beam and electron plasma flow driven by the two-stream instability are in opposite directions. As stated earlier, the radial electric field produced due to two-stream instability has also a defocusing effect on the ion beam as well. The total defocusing force acting on ions is the sum of the electric and magnetic forces and is given by [32, 37]

$$F = e\left(E_x + \frac{v_b}{c}B_y\right) \approx -\frac{1}{4} m_e \nabla_x (v_m^e)^2 - \frac{2\pi e n_b r_b \beta_b^2}{1+\frac{r_b^2 \omega_p^2}{c^2}}\left(1 - \frac{1}{2}\frac{n_p}{n_b}\left(\frac{v_m^e}{v_b}\right)^2\right). \quad (17)$$

Here, we used Eq.(2) for the electric field and Eq.(16) for the self-magnetic field. In the limit when the second term in parenthesis is large compared to the first, Eq.(17) can be simplified to become (for a Gaussian beam profile)

$$F \sim \frac{m_e (v_m^e)^2}{4 r_b}\left(1 + \frac{r_b^2 \omega_p^2}{c^2 + r_b^2 \omega_p^2}\right). \quad (18)$$

Figure 5 compares the Lorentz force in the presence and the absence of the instability. The no-instability case can correspond to for example, propagation in short plasma, such that instability does not have time and space to grow. When comparing the Lorentz forces for the two cases, we first note the difference in the sign of the Lorentz forces. Without instability, the Lorentz force yields beam focusing [16] in contrast to the case with instability. As can be seen from Fig.5 the magnitude of the Lorentz force can be significantly enhanced due to instability.

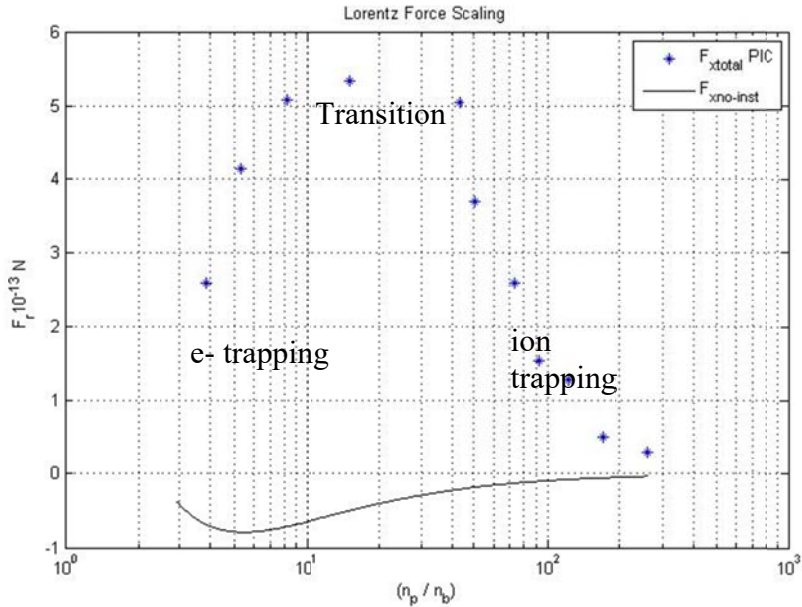

Fig.5 Lorentz Force (radial) $F_x$ as function of $n_p/n_b$. The points correspond to the PIC code results at the time when the instability is saturated; the continuous curve corresponds to the analytical estimate of the total defocusing force for the case without instability.

*The strong dependence of the defocusing Lorentz force acting on beam ions on the background density reveal a potential issue for ballistic propagation of the ion beam in plasma at high plasma density.* Without instability the common assumption is that the higher the plasma density the better is the neutralization [45], because to improve the neutralization, it is intuitive to introduce more background plasma, i.e. increase $n_p$; after all the purpose of the plasma is to reduce the beam space charge and the total current and therefore to reduce the self-fields. However, in the presence of the instability, increasing the plasma density result in increase of the defocusing force until $n_p \sim n_b \left(\frac{m_b}{m_e}\right)^{1/2}$, where the maximum defocusing force is attained.

On the other side using tenuous plasma with the plasma density comparable or even small compared to the beam density can provide good neutralization as was recently demonstrated theoretically[46] and experimentally[47].

## V- Scaling of Defocusing Force with Ion Mass

It is important to note that the saturation mechanism of the two-stream instability and, consequently, the defocusing forces that the beam experiences in plasma also depend on the mass of the ion species, especially in the ion trapping regime. In addition to affecting self-electromagnetic fields, changing the beam ion mass affects the evolution of the transverse beam profile due to the inertia effects. Reference [37] gives estimates at which plasma length the beam defocusing becomes noticeable.

We noted earlier that the scaling parameter, $\alpha$, depends on the ratio of beam ion to electron mass, as well as depends on the ratio of beam and plasma densities. Therefore, it is possible to change the saturation mechanism of the two-stream instability by varying the beam ion mass while keeping the beam and plasma densities constant. In order to study the beam ion mass effect on the saturation mechanism, we simulated a beam-plasma system with fixed beam and plasma densities, $n_b = 2 \times 10^{10}/cm^3$, $n_p = 1.46 \times 10^{12}/cm^3$, respectively and ion mass in the range $m_b/m_e \in [250, 2 \times 10^3]$. We used artificially light ions to better demonstrate the scaling in the ion trapping regime. The ion-beam parameters are $v_b = c/2$ and 12 ns pulse duration as before, but with the beam radius of 5 cm. Figure 6 shows the scaling of electron velocity oscillation amplitude with ion beam mass.

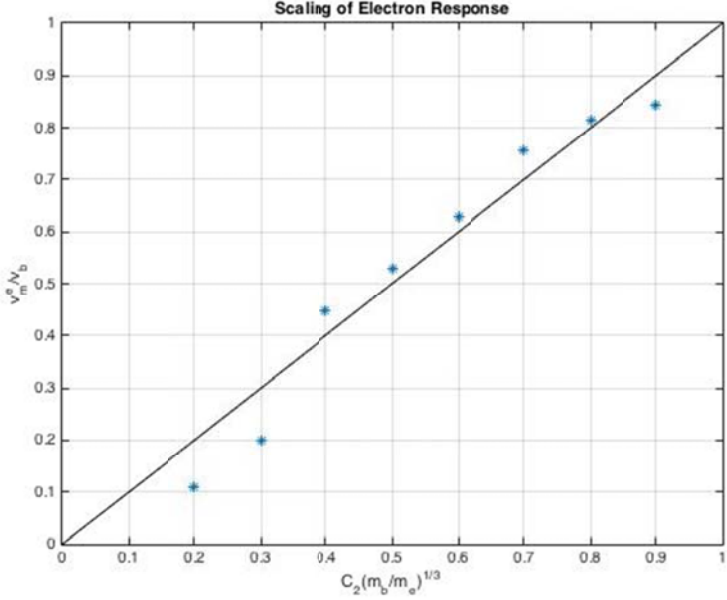

Fig.6 Scaling of the normalized electron velocity oscillation amplitude with the parameter, $\alpha = C_2(m_b/m_e)^{1/3}$, $C_2 = (n_b/n_p)^{2/3}$ for different beam ion mass and $n_b = 2 \times 10^{10}/cm^3$, $n_p = 1.46 \times 10^{12}/cm^3$. The individual points are results from the PIC simulation, the straight line is the theoretical estimate given by Eq.(12). The two-stream instability simulated in all these cases saturates due to the ion trapping mechanism.

Fig.6 shows that in the ion trapping regime, increasing the beam ion mass results in an increase of the amplitude of axial electron oscillation and, consequently, enhances the self-electromagnetic fields, which again will be maximized at the transition region corresponding to $\alpha = 1$.

As discussed earlier in the electron trapping regime, the electron velocity oscillation amplitude reaches the maximum at the beam velocity, $v_m^e = v_b$ and defocusing force does not depend on the ion mass, see Eq.(18).

To estimate effect of the defocusing force on the ion beam, we introduce the characteristic defocusing time, $T_{defocus}$, and the defocusing distance, $L_{defocus}$, which are defined as the time and axial distance that the ion beam propagates before the beam radius doubles due to action of the Lorentz force [24, 37]. In the limit of the beam radius large compared with the skin depth, the Lorentz force is $F \approx m_e(v_m^e)^2/2r_b$ according to Eq. (18); and, in the electron trapping regime, $\approx m_e v_b^2/2r_b$, defocusing force does not depend on ion mass. Correspondingly, the characteristic defocusing time, $T_{defocus}$, and the defocusing distance, $L_{defocus}$, scaling with ion mass are given by [37]:

$$T_{defocus} \sim \left(\frac{r_b}{v_b}\right)\left(\frac{m_b}{m_e}\right)^{\frac{1}{2}}, \quad L_{defocus} = v_b T_{defocus} \sim r_b \left(\frac{m_b}{m_e}\right)^{\frac{1}{2}}.$$

To demonstrate the defocusing effect we simulated two cases with identical beam and plasma parameters but two different ion beam species. In these simulations the background carbon plasma density was $n_p = 2.08 \times 10^{11} cm^{-3}$ and other beam parameters were identical to the ones in Figs.1-4, except for the beam-ion species we used Li$^+$ and K$^+$ instead of H$^+$. Figure 7 shows evolution of the ion beam density profile for both cases. It is evident from Fig.7 that the ion beam density perturbation due to the two-stream instability occurs much faster for the lithium beam than for the potassium beam.

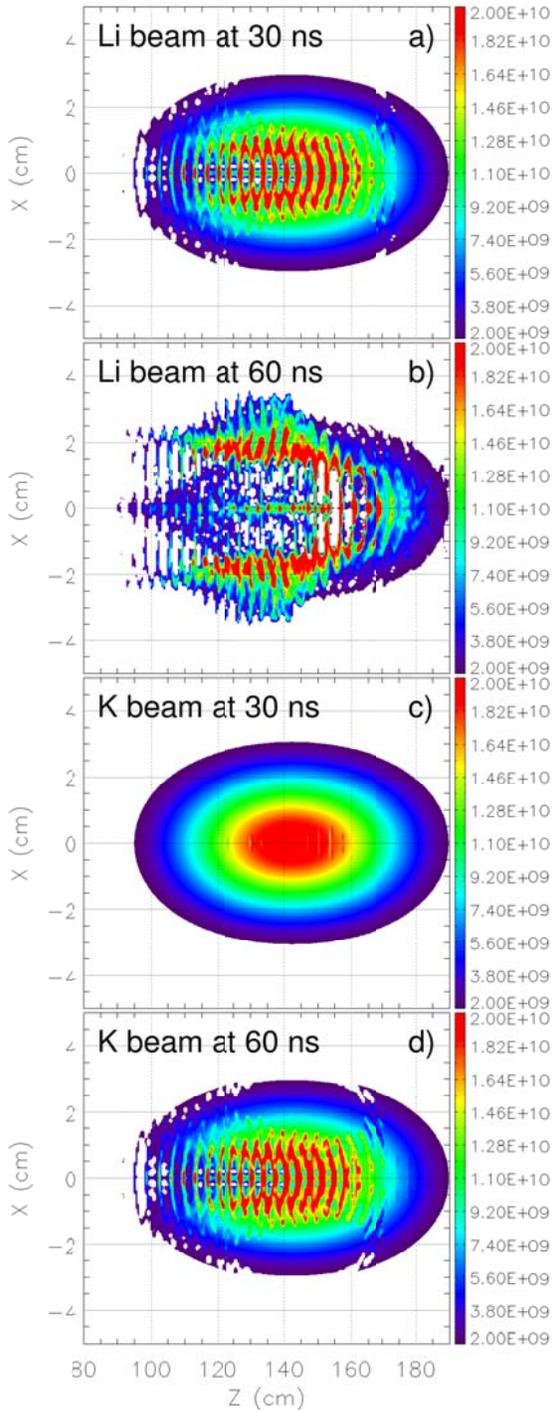

Fig.7 The beam profile evolution for Li⁺ (top) and K⁺ beams (bottom) for time 30 and 60 ns after injection into plasma. The background carbon plasma with $n_p = 2.08 \times 10^{11} cm^{-3}$. Beam parameters are identical to the ones in Fig.1-4, except the beam-ion species are Li⁺ and K⁺ instead of H⁺.

The evolution of the self-magnetic field is shown in Fig.8. The self-magnetic field changes sign after development of instability, compare the magnetic field profile for t=15ns with t=30ns in Fig.8 and Fig.9.

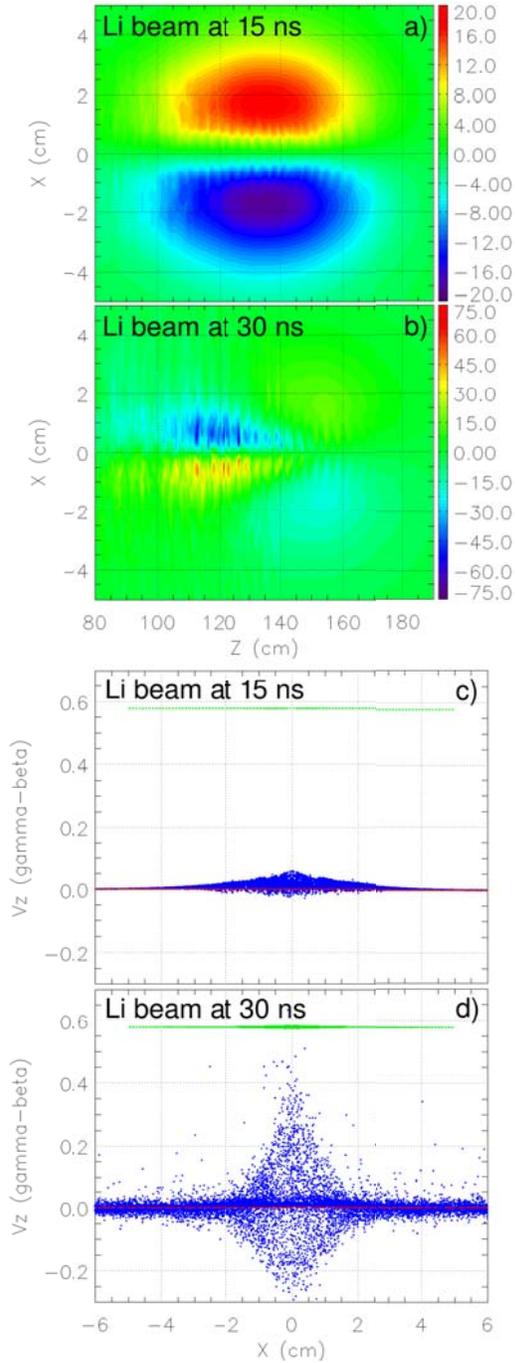

Fig.8 Top: The colorplot of the azimuthal self-magnetic field for Li+ beam 15 ns and 30 ns after injection into the plasma. Bottom: Phase space (axial speed vs. radial location) (green for beam ions and blue for electrons red for C ions). The beam and plasma parameters are the same as in Fig.7.

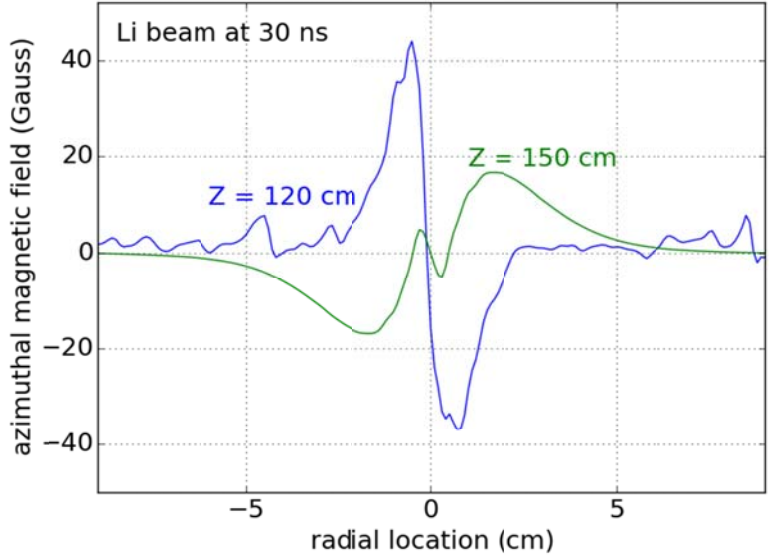

Fig. 9 The radial profile of the azimuthal self-magnetic field for Li+ beam 30 ns after injection into the plasma at two axial locations, (Z = 150 cm, green graph) (Z = 120 cm, blue graph).

## VI- Effects of the Beam Radius Relative to the Skin Depth on the Self-Electric and Magnetic Field at the Saturation of the Two-Stream Instability

We have simulated ion beam transport in background plasma with parameters of the NDCX-II experiment[34]: a Li+ ion beam with the beam velocity, $v_b$ = c/30 corresponding to a directed axial kinetic energy 3.66 MeV, beam density $n_b = 2 \times 10^9 \, cm^{-3}$, that propagates in a background carbon plasma with density $n_p = 1.0 \times 10^{12} cm^{-3}$. The axial beam profile was a Gaussian pulse with pulse width duration Δt=20 ns. The simulations results for the beam radius $r_b = 2.5 cm$ are already reported in Refs.[ 34, 37]. In this section we vary the beam radius from 1mm to 5 cm, which corresponds to transition from the beam radius small to large relative to the skin depth. We study a flat top beam profile, because the ion beam radial profile tends to be flat for high-intensity beams rather than a Gaussian [48]. The beam can be also intentionally transported through an aperture in order to reduce the beam radius [34] and form a flat top beam profile.

Figures 10 and 11 show evolution of the ion beam density and current density after saturation of the two-stream instability, which corresponds to t = 200 ns after beam injection into the plasma or after 2 meters of propagation in plasma. As already discussed in Ref.[34] for these beam and plasma parameters the defocusing forces only affect the ion beam appertured to very small radius of 1mm and do not change beam radial profile for large beam radius. Longitudinal bunching of the ion beam density is high of the order of 100% [ 34, 37] , see Fig. 10. The colorplots of the total current density profiles are shown in Fig.11. Interestingly plasma waves are excited radially even outside of the ion beam pulse at $r > r_b$.

Comparison of the PIC simulation results for the self-magnetic field with the analytical theory given by the Eq.(6) is shown in Fig.12. For flat-top beam radial profile, the self-magnetic field is proportional to the current flowing in the skin layer outside of the beam [16, 45], making use of Eq.(4) this gives

$$B_y = \frac{4\pi n_b v_b}{\omega_p} \left(1 - \frac{1}{2}\frac{n_p}{n_b}\left(\frac{v_m^e}{v_b}\right)^2\right). \tag{19}$$

The value of the generated self-magnetic field does not depend on the beam radius as evident from Fig. 12 and Eq.(19). Similarly, the defocusing force becomes independent of the beam radius for the flat-top beam profile, because the relevant scale of radial inhomogeneity corresponds to the skin depth rather than the beam radius. For the beam parameters shown in Fig.12, $\alpha \sim 1$ and the enhancement of the magnetic field due to instability is a factor of 20. For $r_b = 0.1 cm$, the beam radius is small compared with the skin depth, and self- magnetic field is very small.

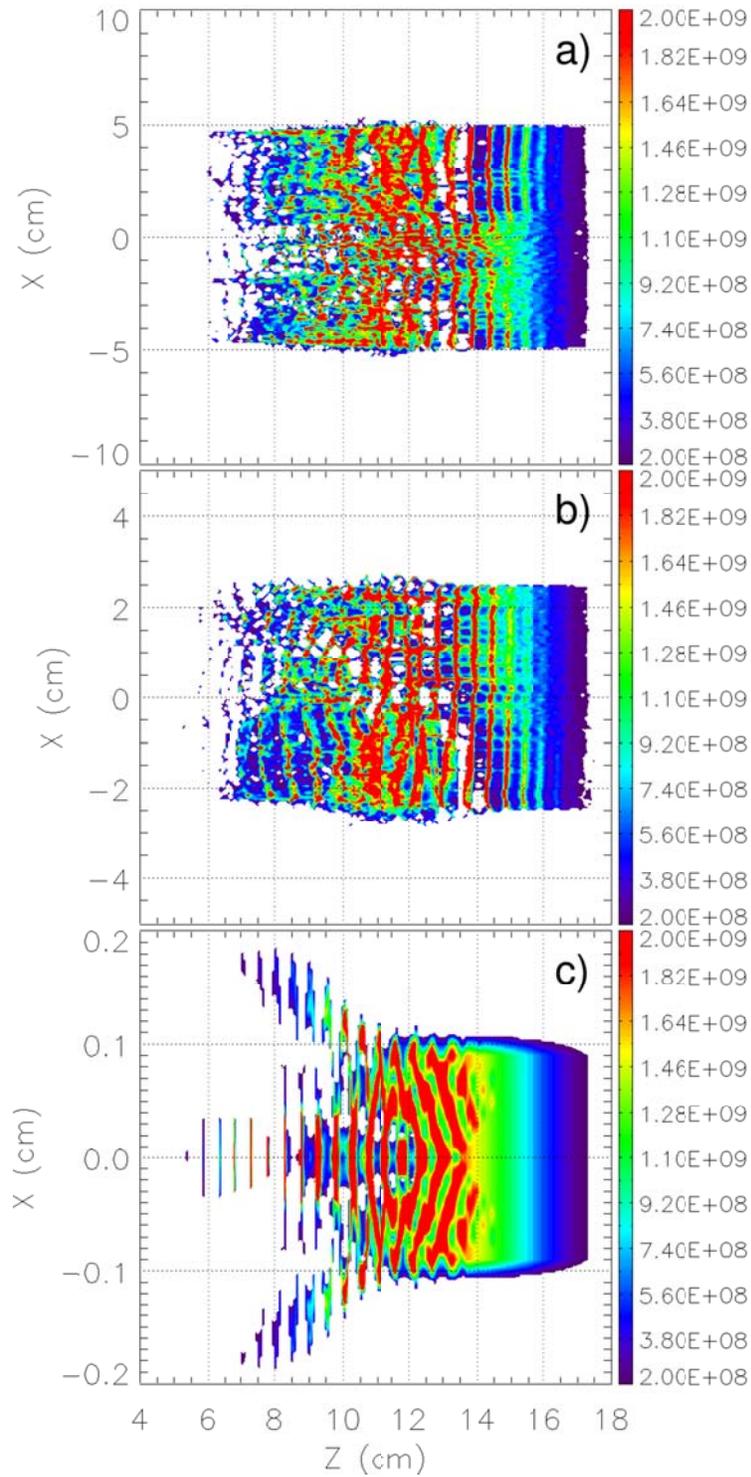

Fig.10 Color plot of the ion beam density profile obtained in PIC simulations for beam and plasma parameters the same as in Fig.1 in Ref. 34 at t = 200 ns after beam injection into the plasma, beam radial profile is flat top (with rounded edges) with three values of the beam radius (a) $r_b = 5cm$ (b) $r_b = 2cm$ (c) $r_b = 0.1cm$

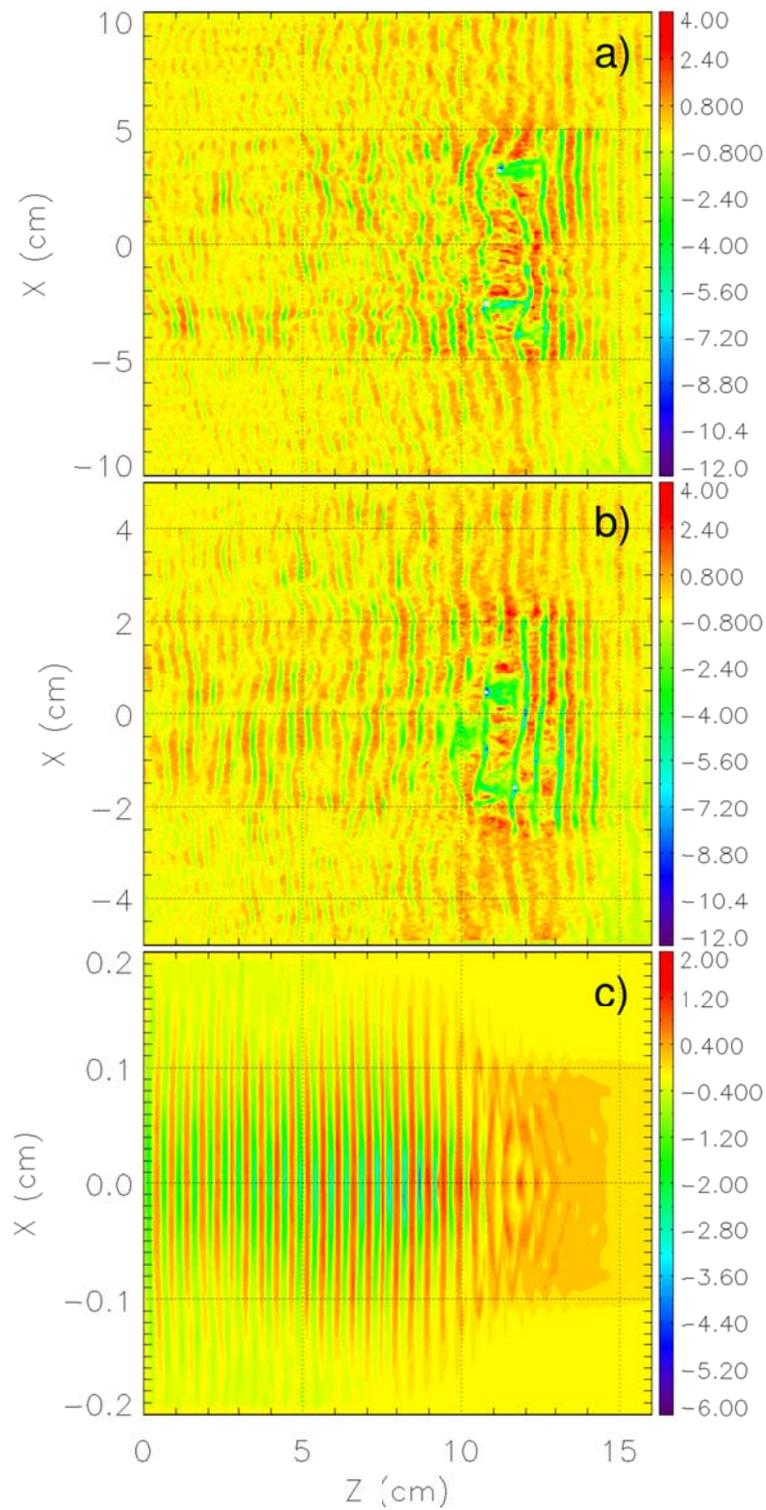

Fig. 11 Colorplot of the axial current density for same simulation parameters identical to Fig. 10. Current is in A/cm$^2$.

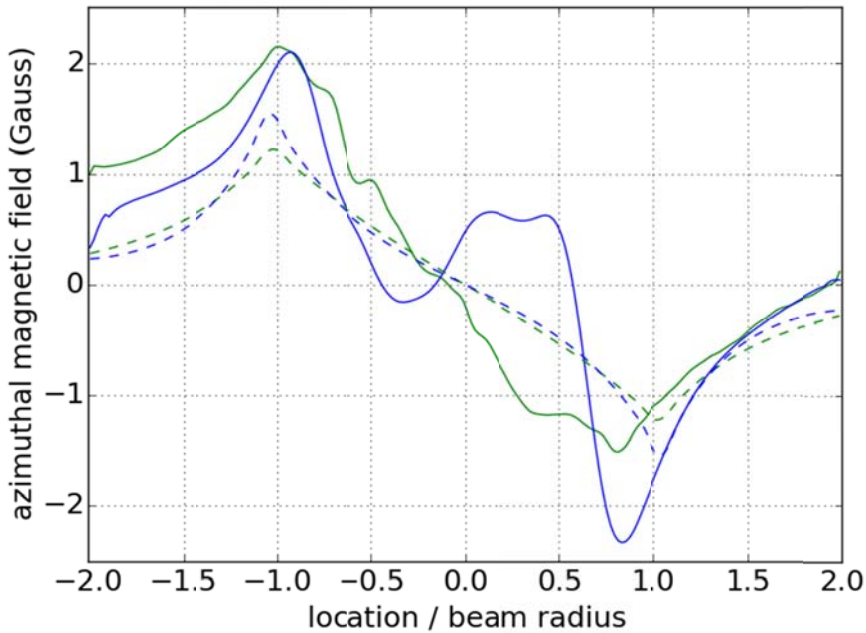

Fig. 12 Radial profile of the self-magnetic field obtained in the PIC simulation (solid lines) with the analytical theory given by the Eq.(6) (dashed lines) are shown for the same conditions as in Figs. 10, 11 for cases (a) $r_b = 5cm$ (blue) (b) $r_b = 2cm$ (green), for case (c) $r_b = 0.1cm$ the self-magnetic field is much smaller.

Note that our simulations do not show any development of electromagnetic instabilities, neither transverse two-stream nor hose instability[49,50,51]. The hose instability does not have time to develop, because the beam is not relativistic. The transverse two-stream instability can develop for the two electron streams with a discontinuous radial profile [49,50]. However, for the self-consistent profile of the return current that is a smooth radial function on the distances of order of the skin depth the instability growth rate is strongly reduced. It was also shown in Ref. [52] that for a finite transverse beam size there are no eigenmodes associated with the two-stream instability. Interestingly, if the self-magnetic field of the beam is also taken into account, the transverse two-stream instability does not develop at all[53]. Similar conclusions were drawn in experimental study of Ref. 28. Our particle-in-cell simulation results also confirm that.

## VII. Two-Stream Instability for a Proton Beam Generated by a Laser.

In this section, we simulate another ion beam pulse interaction with background plasma. The beam and plasma parameters correspond to experiments performed at GSI, Helmholtz center for heavy ion research. [54] A proton beam with the peak ion beam energy of 7.8 MeV was reported in Ref [54]. This corresponds to approximately $v_b = (4/30)\,c$. After compression of the beam via focusing, the observed pulse length was $\tau = (462 \pm 40)$ ps, the peak particle current was 170 mA, and the minimum transverse beam size at the longitudinal focus position was measured to $3 \times 18 \, \text{mm}^2$, resulting in approximately 0.32 A/cm² of ion beam current density.

In order to demonstrate the possible effect of the two-stream instability for future experiments with proton beams generated by laser, we employ the following assumptions: (1) the pulse length is larger than the observed value and (2) the background plasma density is chosen to be two or three orders larger than the ion beam density, so that the plasma wave with multiple wave lengths ($L = 2\pi v_b/\omega_p$) can develop due to two-stream instability. Hence, we chose for beam pulse duration $t_{pulse} = 9 \, ns$, and for plasma density, $n_p = 5.0 \times 10^{11} cm^{-3}$. A moving window algorithm was used: the moving window starts 11 ns after the ion beam injection and the speed of the moving window is approximately $3.8 \times 10^9$ cm/s, which is slightly slower than the ion beam velocity $v_b = 4 \times 10^9$ cm/s. This is due to the phase velocity of the plasma wave being smaller than the ion beam velocity as shown in Eq. (10).

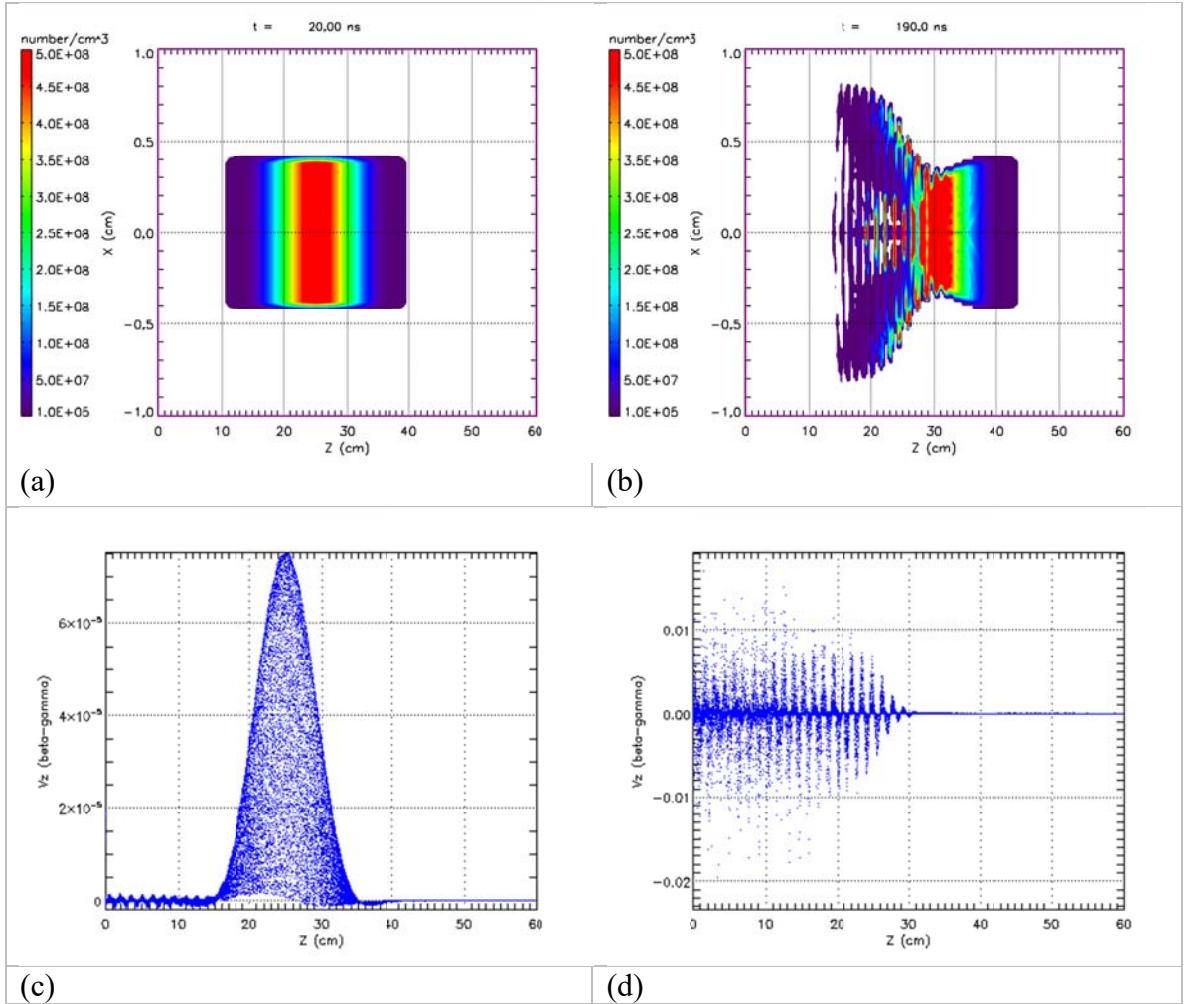

Fig. 13. The two-dimensional PIC simulation in Cartesian coordinate for ion beam and plasma parameters: $n_p = 5.0 \times 10^{11} cm^{-3}$, $n_b = 5 \times 10^8 \, cm^{-3}$, $r_b = 4 \, cm$, total $t_{pulse} = 9 \, ns$, $v_b/c = 4/30$. Shown are the ion beam density colorplots: prior to instability (t = 20 ns) Fig.13(a) and after 7.6 m of propagation (t = 190 ns) in plasma when instability develops and distorts the initial beam profile Fig.13 (b). Figures

13(c),(d) show the electron distribution in the phase space in the axial direction (Z vs. V$_z$) at the same times.

Figure 13 shows the electron phase space in the axial direction (Z vs. V$_z$) as well as the ion beam density profiles before and after the two-stream instability develops. The initial beam profile is a Gaussian pulse in the longitudinal direction and has a flattop distribution in the transverse direction in a slab geometry. Note that the electrons are slightly accelerated initially to neutralize the charge and current of the ion beam $\frac{v}{c} = \frac{n_b}{n_b+n_p}\frac{v_b}{c} \approx 7.5 \times 10^{-5}$, which can also be seen in the simulation results. As the instability develops, the electrons are perturbed and plasma waves are generated. From Eq. (13), $\alpha = (1 \times 10^{-3})^{2/3} \, 1836^{1/3} = 0.122 < 1$, therefore the saturation mechanism of the instability is due to ion beam trapping. It can be seen from Fig 13(d) that electron acceleration is moderate and the maximum electron velocity is an order of magnitude smaller than the ion beam velocity. The simulation results show that the two-stream instability can play an important role for the beam quality if the pulse length is long and if the background plasma density is large enough that the instability can develop within the ion beam pulse.

## VIII-Conclusion

In this review, we described the effects of beam-driven two-stream instability on propagation of the ion beam pulse in background plasma. The self-electromagnetic field generated by the ion beam pulse during propagation in plasma before the two-stream instability develops focuses the ion beam. In contrast to this, we showed that the non-linear electromagnetic fields generated by the two-stream instability can result in significant defocusing of the beam.

The magnitude of the self-electromagnetic fields depends strongly on the saturation amplitude of axial electron velocity oscillations. By identifying a scaling parameter, $\alpha = \left(\frac{n_b}{n_p}\right)^{\frac{2}{3}} \left(\frac{m_b}{m_e}\right)^{1/3}$, which is a function of the ratio of the beam to plasma densities, and beam ion to electron masses, we studied the scaling of the non-linear self-electromagnetic fields in wide range of the plasma densities and ion masses. We showed that in the limit of low plasma density, $n_p < n_b \left(\frac{m_b}{m_e}\right)^{1/2}$, the instability saturates by the electron trapping mechanism. In the opposite limit, $n_p > n_b \left(\frac{m_b}{m_e}\right)^{1/2}$, the instability saturates by the ion beam trapping mechanism. The azimuthal self-magnetic field and the total defocusing force have a maximum in the transition region, $n_p \sim n_b \left(\frac{m_b}{m_e}\right)^{1/2}$. We identified this transition region as the least favorable for a neutralized ballistic propagation of the ion beam in background plasma due to deleterious effects of the two-stream instability. We also showed that in the electron trapping regime ($n_p < n_b \left(\frac{m_b}{m_e}\right)^{1/2}$) increasing the plasma density causes an increase of the total defocusing force and can strongly affect ballistic propagation in background plasma. This finding is in contrast to previous neutralization studies where the effect of the two stream instability was not taken into account and it was assumed that denser plasma is better for neutralization of the ion beam pulse.

This research was funded by the U.S. Department of Energy, Office of Fusion Energy Sciences.